\documentclass[final,english]{bullsrsl}[2022/06/15]

\usepackage[latin1]{inputenc}
\usepackage[T1]{fontenc}

\usepackage{natbib} 
\usepackage{graphicx}

\begin{document}
\title{Extraction of YSO Cores and Active Regions near Star-forming Site AFGL 5157}

\author[affil={1}, corresponding]{Aayushi}{Verma}
\author[affil={1}]{Saurabh}{Sharma}
\author[affil={2}]{Lokesh}{Dewangan}
\affiliation[1]{Aryabhatta Research Institute of observational sciencES (ARIES), Manora Peak, Nainital-263001, India}
\affiliation[2]{Physical Research Laboratory, Navrangpura, Ahmedabad-380 009, India}
\correspondance{aayushiverma@aries.res.in}
\date{20th May 2023}
\maketitle

% \author[affil1]{FirstName (+ MiddleInitials if necessary)}{FamilyName}
% \author[affil2]{...}{}
% \equalcontribauthor[]{}{} % Maximum two --> counter
% \consortium[affil]{Consortium Name}
% With consortium: affiliation will be set to "See Appendix 1 for a full
% list of consortium members and their respective affiliations
% \affiliation[affil1]{...}
% \affiliationq[affil2]{...}

% \correspondence[]{}
% No explicit corresponding author: use first author
% 

% Abstract of the paper in the same language as the paper
\begin{abstract}
We have carried out a quantitative analysis of the $1^{\circ} \times 1^{\circ}$ region near star-forming site AFGL 5157 using \emph{'Minimal Spanning Tree'} (MST). The analysis reveals that this region consists of five major clusters. The cluster radii of the cores and active regions were found to be varying between 0.75-2.62 pc and 2.77-4.58 pc, respectively, for these regions, while the aspect ratio varies between 0.71 to 7.17. This hints towards the clumpy as well as elongated clusters in the region. We calculated structure parameter $Q$ for each region which varies between 0.41-0.62 and 0.23-0.81 for the cores and ARs, respectively. This shows the existence of fractal distribution in all the cores and ARs except the core of the [HKS2019] E70 bubble.
\end{abstract}

\keywords{Interstellar medium, Star formation, Star-forming regions}

%\section{Section -- Level 1 title (Times New Roman, bold, 14 pts)}

\section{Introduction}
The formation of stars occurs in a group of clusters and associations, and it is assumed that they cannot be formed in isolation \citep{2003ARA&A..41...57L}. Observational analyses of embedded star-forming regions reveal that the distribution of stars is generally elongated, clumpy, or both (\citealt{2008ApJ...675..491A,2008ApJ...688.1142K}). This distribution is correlated with the distribution of dense molecular gas of natal clouds \citep{2008ApJ...675..491A}. The degree to which a cloud can form a group or association of stars is majorly governed by the dense material in the molecular cloud \citep{2003ARA&A..41...57L}. The star formation scenario and the physical processes affecting the star formation in the region can be well understood by the mapping of young stellar objects (YSOs; \citealt{2008ApJ...688.1142K}). Various surveys of molecular clouds reveal that approximately 75 \% of the embedded young stars exist in groups their clusters have equal to or more than 10 members \citep{1993prpl.conf..429Z,2007prpl.conf..361A}. The quantitative structural analysis of these clusters; the sizes, densities, and morphologies of young stellar cores: can be used to examine the theoretical models of star formation \citep{2014ApJ...787..107K}.

\section{Source Selection}
Active star-forming regions in molecular clouds generally consist of young star clusters (YSCs), mid-infrared (MIR) bubbles, clouds/filaments, and massive stars.
Previous studies have suggested that the expansion of bubbles associated with the H\,{\sc ii} regions can trigger 14 to 30\% of the formation of the stars (e.g., \citealt{2010A&A...523A...6D}, \citealt{2012ApJ...755...71K}).

For the present study, we have selected a poorly explored region of $1^{\circ} \times 1^{\circ}$ which consists of five clusters: AFGL 5157 ($\alpha=05^h37^m48^s$ and $\delta=+31^{\circ}59^{'}44^{''}$), MIR bubble [HKS2019] E70 ($\alpha=05^h38^m19^s$ and $\delta=+31^{\circ}44^{'}22^{''}$), [FSR2007] 0807 ($\alpha=05^h36^m39^s$ and $\delta=+31^{\circ}49^{'}20^{''}$), [KPS2012] MWSC 0620 ($\alpha=05^h39^m20^s$ and $\delta=+31^{\circ}30^{'}58^{''}$) and an H\,{\sc ii} region IRAS 05221+3115 ($\alpha=05^h36^m26^s$ and $\delta=+31^{\circ}17^{'}43^{''}$).  

\begin{figure}
\centering
\includegraphics[scale=0.52]{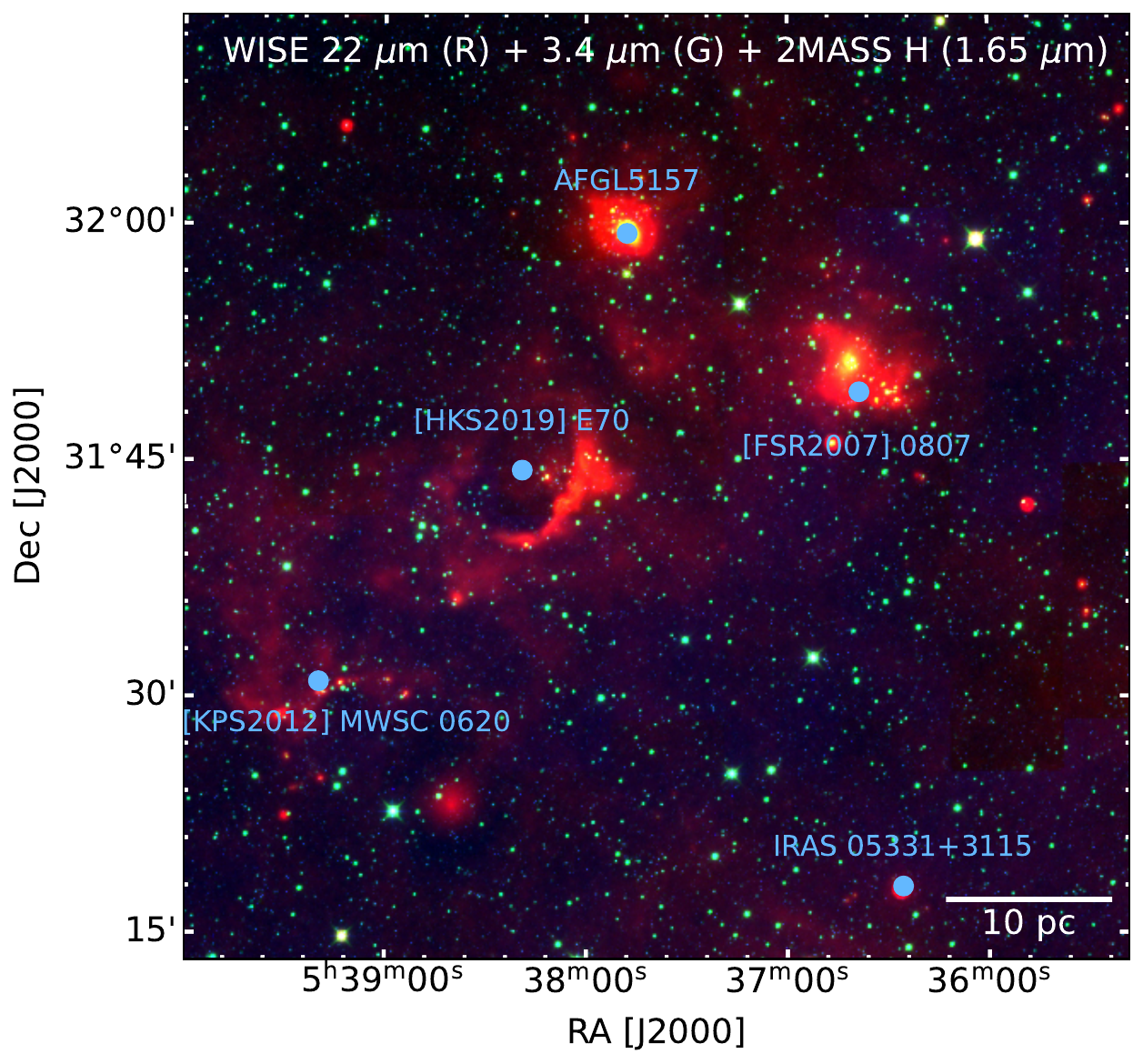}
\bigskip
\begin{minipage}{12cm}
\caption{Color-composite image (Red: \emph{WISE} 22 $\mu$m; Green: \emph{WISE} 3.4 $\mu$m; Blue: 2MASS H band, 1.65 $\mu$m) of $1^{\circ} \times 1^{\circ}$ region overlaid with the locations of five clusters (blue circle).}
\label{fig:wise22}
\end{minipage}
\end{figure}

Figure \ref{fig:wise22} shows the color-composite image (Red: \emph{WISE} 22 $\mu$m; Green: \emph{WISE} 3.4 $\mu$m; Blue: 2MASS H band, 1.65 $\mu$m) of $1^{\circ} \times 1^{\circ}$ region overlaid with the locations of five clusters (blue color). The \emph{WISE} 22 $\mu$m represents the distribution of warm gas. The MIR emission in 3.4 $\mu$m wavelength indicates the distribution of gas and dust. All these features coincide with the locations of identified clusters indicating strong evidence of star formation activities.

% Figure \ref{fig:wise22} represents the \emph{WISE} 22 $\mu$m image of this region which represents the distribution of warm gas. This image is overlaid with the locations of five clusters.

% \begin{figure}
%     \centering
%     \includegraphics[width=0.75\textwidth]{rgb_1deg_bina.pdf}
%     \caption{Color-composite image (Red: \emph{WISE} 22 $\mu$m; Green: \emph{WISE} 3.4 $\mu$m; Blue: 2MASS H band, 1.65 $\mu$m) of $1^{\circ} \times 1^{\circ}$ region overlaid with the locations of five clusters (blue circle).}
%     \label{fig:wise22}
% \end{figure}

\section{Extraction of the Cores and the Active Regions (ARs)}

% Footnotes or endnotes are \emph{not allowed}. 

\begin{figure}
\centering
    \includegraphics[scale=0.3]{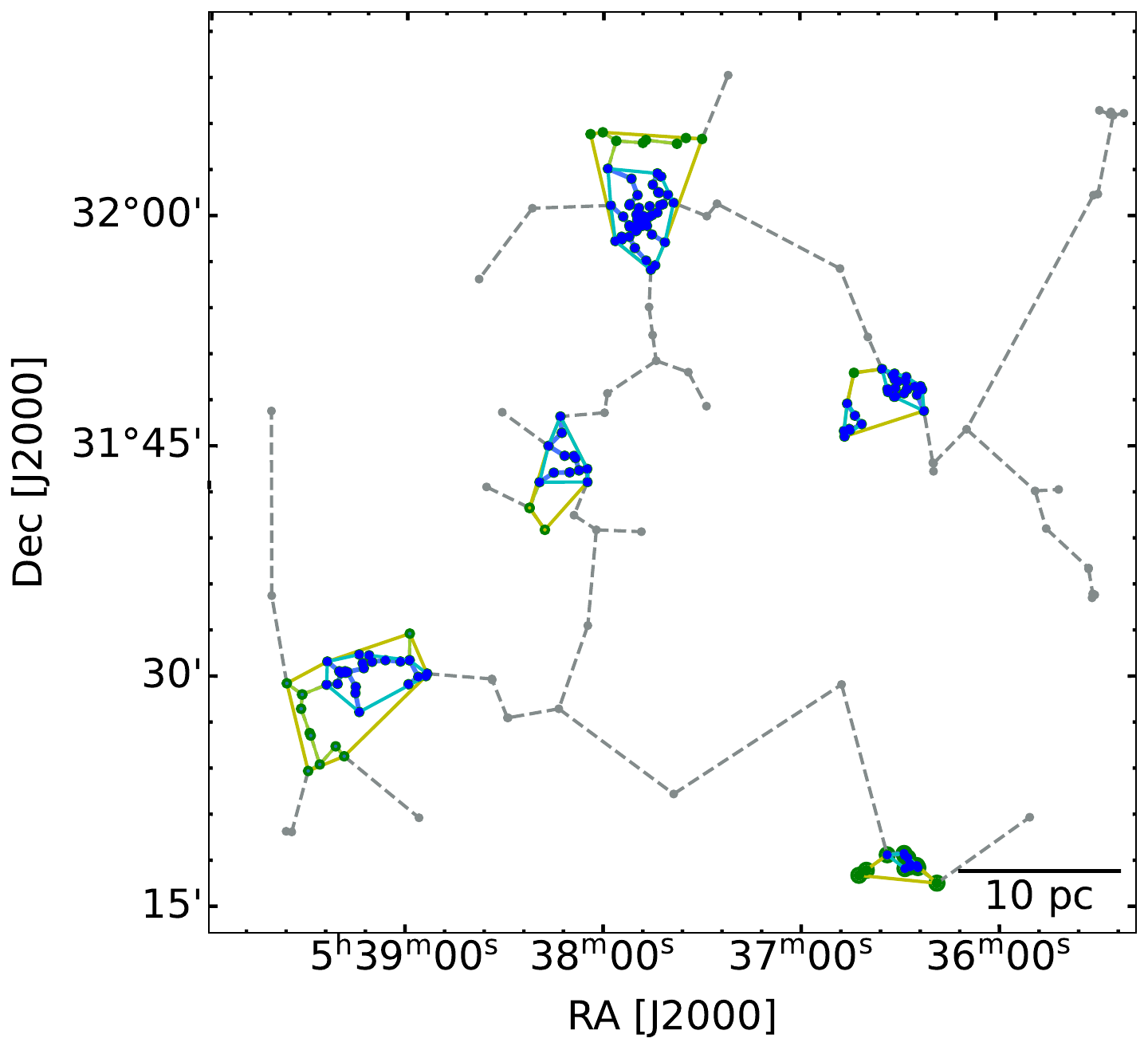}
    \includegraphics[scale=0.365]{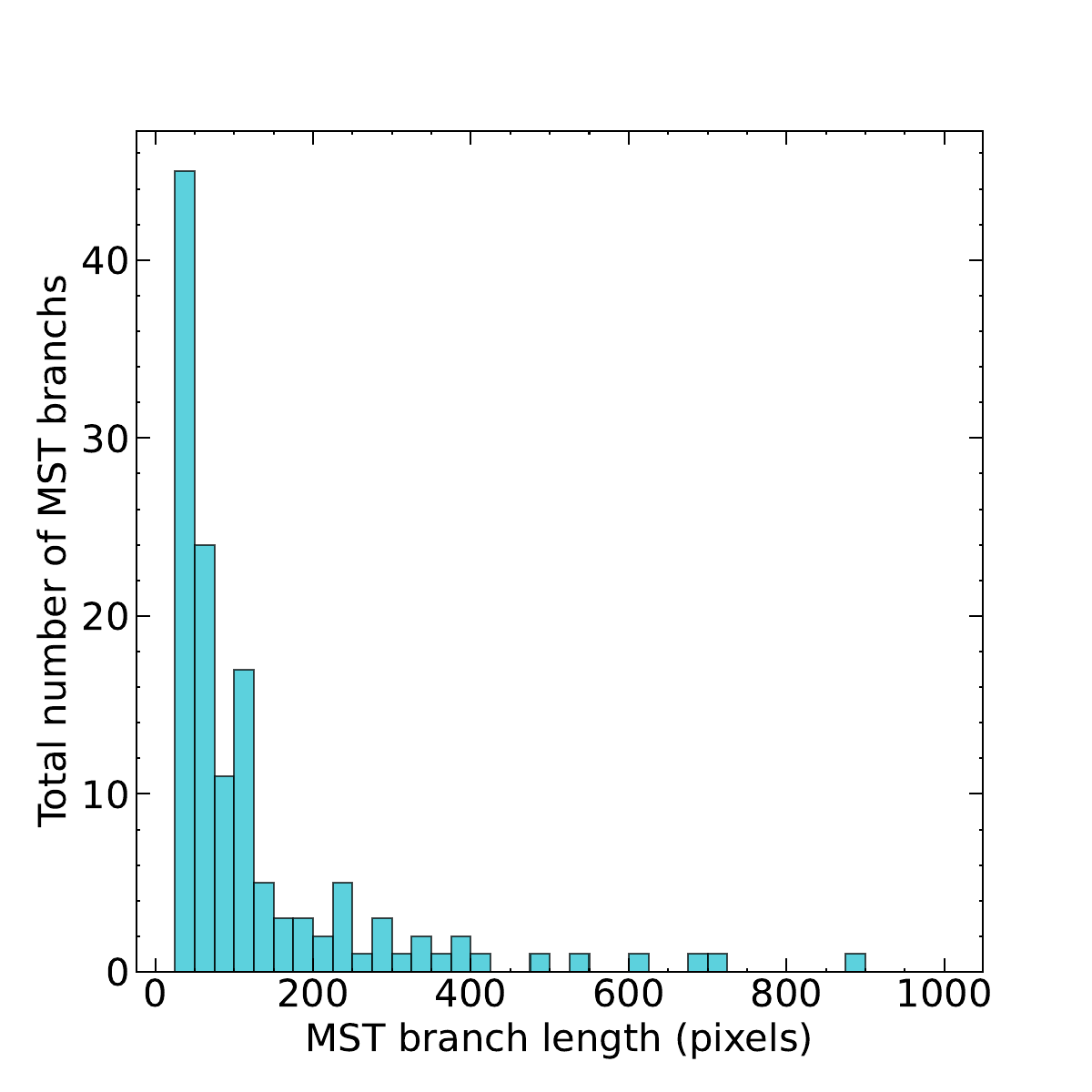}
\bigskip
\begin{minipage}{12cm}
\caption{Left panel: MST connections in the core (blue) and AR (green) and the locations of YSOs (blue dots for the YSOs in the core and green dots for the YSOs in AR). The isolated cores and ARs are also enclosed by the \emph{Convex hulls} using cyan and yellow lines, respectively. The locations of the identified YSOs are also marked by black dots. Right panel: Histogram of the MST branch lengths.}
\label{fig:mst}
\end{minipage}
\end{figure}

Assuming that the five clusters in the region might be fragmented from the same molecular cloud, we isolated them by applying an empirical technique \emph{'Minimal Spanning Tree'} (MST; \citealt{2009ApJS..184...18G}).
This technique finds edges with minimum weight in each iteration and then adds them to the MST. The advantage of this technique is that it isolates the subgroups without biasing or smoothening and prevents basic geometry \citep{2004MNRAS.348..589C,2009ApJS..184...18G, 2020ApJ...891...81P}. We used this technique on the identified YSOs based on excess IR emission. The MST is plotted in the left panel of Figure \ref{fig:mst}, which points towards five subgroups in the region having different concentrations of YSOs. The dots and lines in different colors are representing the YSOs and the MST branches, respectively. We chose a surface density threshold stated as critical branch length to isolate the subgroups by plotting a histogram between MST branch lengths and MST branch numbers (right panel of Figure \ref{fig:mst}). The histogram peaks at a relatively small spacing than large spacing which indicates the existence of significant subgroups. If the sources have a branch length less than the critical branch length, then they are considered part of the same cluster.

% \begin{figure}
%     \centering
%     \includegraphics[width=0.48\textwidth]{mst_Qhull_icrs_1deg_bina.pdf}
%     \includegraphics[width=0.50\textwidth]{nodes_distances_hist_bina.pdf}
%     \caption{Left panel: MST connections in the core (blue) and AR (green) and the locations of YSOs (blue dots for the YSOs in the core and green dots for the YSOs in AR). The isolated cores and ARs are also enclosed by the \emph{Convex hulls} using cyan and yellow lines, respectively. The locations of the identified YSOs are also marked by black dots. Right panel: Histogram of the MST branch lengths.}
%     \label{fig:mst}
% \end{figure}

These sub-groups are enclosed by selecting a point where the shallow-sloped segment has a gap in the distribution of MST branch lengths. These regions are termed as 'active regions' (AR). It is considered that these regions have moved out of the core (overdense regions) because of dynamic evolution. The critical branch lengths are 100  and 125 pixels for the cores and ARs, respectively. The YSOs and MST connections in the cores and ARs are represented by blue and green dots and lines, respectively, in the left panel of Figure \ref{fig:mst}. 
The isolated cores and ARs are enclosed by the \emph{Convex hulls} using cyan and yellow lines, respectively. All the clusters are found to have a single core, whereas [FSR2007] 0807 is found to have two cores which are termed C1 and C2.
The locations of all the identified YSOs are shown by black dots. 
We have also calculated the radius of the cores and ARs ($R_{cluster}$), $R_{circ}$, 
% (see section \ref{sec:clustering})
and aspect ratio. $R_{circ}$ is expressed as half of the distance between the two most distant hull points, whereas the aspect ratio is $\frac{R^2_{circ}}{R^2_{cluster}}$. The evaluated parameters of identified regions are given in Table \ref{tab:mst_parameter}.

The shape of the cluster may not always be circular. So, the cluster area is redefined by \emph{convex hull}, which is a polygon enclosing all points of a grouping with internal angles less than 180$^{\circ}$ between two contiguous sides (\citealt{2006A&A...449..151S}, \citealt{2020MNRAS.498.2309S}). The \emph{convex hull} is used to evaluate the area of the cluster $A_{cluster}$ \citep{2020MNRAS.498.2309S}, given as:

\begin{equation}
        A_{cluster}=\frac{A_{hull}}{1-\frac{n_{hull}}{n_{total}}}
\end{equation}
where $A_{hull}$ is the hull area, $n_{hull}$ is the total number of hull vertices and $n_{total}$ is the number of points lying inside the hull. $R_{cluster}$ is the radius of the circle whose area is equal to $A_{cluster}$.

\begin{table*}[t]
    \centering
    \caption{Properties of the Identified Cores and ARs}
    \begin{tabular}{c c c c c c c c c}
    \hline
    Region & N$^a$ & N$^b$ & N$^c$ & V$^d$ & $R_{cluster}$ & $R_{circ}$ & Aspect & $Q$\\
     & & &  &  & (pc) & (pc) & Ratio & parameter\\
    \hline
    For Cores\\
    AFGL5157 & 52 & 2 & 50 & 9 & 2.62 & 3.39 & 3.96 & 0.62\\
    $[$HKS2019$]$ E70 & 12 & 0 & 12 & 5 & 2.59 & 2.19 & 0.71 & 0.50\\
    $[$FSR2007$]$ 0807 (C1) & 16 & 4 & 20 & 7 & 1.29 & 1.82 & 2.00 & 0.55\\
    $[$FSR2007$]$ 0807 (C2) & 7 & 2 & 5 & 4 & 0.89 & 1.02 & 1.31 & 0.44\\
    $[$KPS2012$]$ MWSC 0620 & 25 & 5 & 20 & 9 & 2.61 & 3.10 & 1.41 & 0.39\\
    IRAS 05221+115 & 10 & 0 & 10 & 5 & 0.75 & 2.00 & 7.17 & 0.24\\
    \hline\\
    For ARs\\
    AFGL5157 & 60 & 2 & 58 & 7 & 3.57 & 4.58 & 1.64 & 0.38\\
    $[$HKS2019$]$ E70 & 15 & 1 & 14 & 6 & 2.59 & 3.53 & 1.86 & 0.81\\
    $[$FSR2007$]$ 0807 & 30 & 5 & 25 & 9 & 2.56 & 2.80 & 1.20 & 0.61\\
    $[$KPS2012$]$ MWSC 0620 & 36 & 10 & 26 & 7 & 3.95 & 5.27 & 1.78 & 0.37\\
    IRAS 05221+115 & 13 & 3 & 9 & 4 & 1.44 & 2.77 & 3.70 & 0.23\\
    \hline\\
    \end{tabular}
    
    Notes: The central position of the sub-regions along with the number of enclosed YSOs are mentioned in column 2,3,4; the number of vertices of the convex hull, radius of the cluster, circle radius, and the aspect ratio are given in columns 5, 6, 7, and 8, respectively.
    
    $^a$Number of enclosed YSOs\\
    $^b$Number of enclosed Class $\textsc{i}$ YSOs\\
    $^c$Number of enclosed Class $\textsc{ii}$ YSOs\\
    $^d$Number of the vertices of the convex hull\\
    
    \label{tab:mst_parameter}
\end{table*}

\section{Result and Conclusion}
We examined the spatial distribution of the YSOs in the $1^{\circ} \times 1^{\circ}$ region near the star-forming site AFGL 5157. MST analysis of the region reveals the presence of five major clustering in the region. We have determined the basic structural parameters of these regions and found that the radius of the cluster varies between  0.75 pc to 2.62 pc with a mean value of 1.74 pc. In contrast, the aspect ratio varies between  0.71 to 7.17. Thus this region consists of clumpy as well as elongated clusters. We also calculated structure parameter $Q$ for each region (see Table \ref{tab:mst_parameter}) and found that it varies between 0.41-0.62 and 0.23-0.81 for the cores and ARs, respectively. The large $Q$ values (i.e., $>$ 0.8) are related to the centrally condensed spatial distributions, whereas small $Q$ values (i.e., $<$ 0.8) are related to the fractal substructures \citep{2019A&A...629A.135D}. This shows the existence of fractal distribution in all the cores and ARs except the core of [HKS2019] E70. Since [HKS2019] E70 has $Q>0.8$, thus it has a centrally condensed spatial distribution.

\begin{acknowledgments}
This publication makes use of data from the Two Micron All Sky Survey, which is a joint project of the University of Massachusetts and the Infrared Processing and Analysis Center/California Institute of Technology, funded by the National Aeronautics and Space Administration and the National Science Foundation. This work is based on observations made with the \emph{Spitzer} Space Telescope, which is operated by the Jet Propulsion Laboratory, California Institute of Technology under a contract with the National Aeronautics and Space Administration. This publication makes use of data products from the Wide-field Infrared Survey Explorer, which is a joint project of the University of California, Los Angeles, and the Jet Propulsion Laboratory/California Institute of Technology, funded by the National Aeronautics and Space Administration. A.V. acknowledges the financial support of DST-INSPIRE (No.$\colon$ DST/INSPIRE Fellowship/2019/IF190550).
\end{acknowledgments}

\begin{furtherinformation}

\begin{orcids}
\orcid{0000-0002-6586-936X}{Aayushi}{Verma}
\orcid{0000-0001-5731-3057}{Saurabh}{Sharma}
\orcid{0000-0001-6725-0483}{Lokesh}{Dewangan}
\end{orcids}

\begin{authorcontributions}
% This section is mandatory when there is more than one author.
% The contributions of each author (identified by their initials) must be declared.
% We recommend to follow the \href{http://credit.niso.org}{CRediT} taxonomy (Contributor Roles Taxonomy).
The present study results from a collaboration to which all the authors have significantly contributed.
\end{authorcontributions}

\begin{conflictsofinterest}
% This section is \emph{mandatory}. Authors must declare any personal or professional circumstances that may be perceived as influencing the research reported in the paper. If there is no conflict of interest, please state that ``
The authors declare no conflict of interest.
\end{conflictsofinterest}

\end{furtherinformation}

\bibliographystyle{bullsrsl-en}

\bibliography{bibliography}

\end{document}